%% file: IEEE_Microwave_QC_Vertical.tex
\documentclass[conference,10pt,letterpaper]{IEEEtran}%
\usepackage{amsmath}
\usepackage{times}
\usepackage{graphicx}
\usepackage{multirow}
\usepackage[none]{hyphenat}
\usepackage{float}
\usepackage{subfig}
\usepackage{siunitx}
\usepackage{chemformula}
\usepackage{braket}
\usepackage{amssymb}
\input{IMS_modify_IEEEtran_18b_CTAN_V3c}

\begin{document}
\raggedbottom

\title{Microwave Packaging for Superconducting Qubits}
\IMSthispaperforblindreview

\IMSthispaperforfinalpublication
\IMSauthor{\IMSauthorblockNAME{
Benjamin Lienhard\IMSauthorrefmark{1},
Jochen Braum\"uller\IMSauthorrefmark{2},
Wayne Woods\IMSauthorrefmark{3},
Danna Rosenberg\IMSauthorrefmark{3},
Greg Calusine\IMSauthorrefmark{3},\\
Steven Weber\IMSauthorrefmark{3},
Antti Veps\"al\"ainen\IMSauthorrefmark{2},
Kevin O'Brien\IMSauthorrefmark{1,2},
Terry P. Orlando\IMSauthorrefmark{1,2},
Simon Gustavsson\IMSauthorrefmark{2},\\
William D. Oliver\IMSauthorrefmark{2,3,4,*},
}
\\%
\IMSauthorblockAFFIL{
\IMSauthorrefmark{1} Department of Electrical Engineering \& Computer Science, Massachusetts Institute of Technology, USA\\
\IMSauthorrefmark{2}Research Laboratory of Electronics, Massachusetts Institute of Technology, USA\\
\IMSauthorrefmark{3}MIT Lincoln Laboratory, USA\\
\IMSauthorrefmark{4}Department of Physics, Massachusetts Institute of Technology, USA
}
\\%
\IMSauthorblockEMAIL{
\IMSauthorrefmark{*}william.oliver@mit.edu
}
}

\maketitle
%
%
%
\begin{abstract}
Over the past two decades, the performance of superconducting quantum circuits has tremendously improved. The progress of superconducting qubits enabled a new industry branch to emerge from global technology enterprises to quantum computing startups. Here, an overview of superconducting quantum circuit microwave control is presented. Furthermore, we discuss one of the persistent engineering challenges in the field\hspace{0.03cm}---\hspace{0.03cm}how to control the electromagnetic environment of increasingly complex superconducting circuits such that they are simultaneously protected and efficiently controllable.
\end{abstract}
%

\section{Introduction}

Universal quantum computation aims to perform specific computational problems such as integer factorization \cite{Shor1994}\hspace{0.03cm}---\hspace{0.03cm}central to cryptography protocols\hspace{0.03cm}---\hspace{0.03cm}or database search \cite{Grover1996} in a significantly more efficient way than classical computers. The superconducting qubit modality is a leading candidate today for the realization of such a quantum information processor. Over the last 15 years, the fabrication, design, and control of superconducting qubits have considerably improved, resulting in exponential advancements of coherence properties \cite{Oliver2013}. The increase in qubit performance has enabled the demonstration of several major milestones in the pursuit of scalable quantum computation. Among others, multi-qubit control and entanglement techniques \cite{Steffen2006,Barends2014}, improved quantum gate fidelities \cite{Gustavsson2013}, and better readout schemes \cite{Blais2004,Wallraff2004} have enabled the demonstration of small-scale quantum algorithms \cite{Lucero2012, Colless2018}. 
However, many engineering challenges need to be overcome to realize the full promise of quantum computation. Here, we present a brief overview of superconducting circuits and present simulations of a sample package with the goal to simultaneously achieve efficient qubit control while suppressing qubit energy decay channels. 



\section{Experimental Methods} \label{sec: background and basics}
\subsection{The Transmon Qubit}

\begin{figure*}[ht!]
\centering
\includegraphics[width=0.96\textwidth]{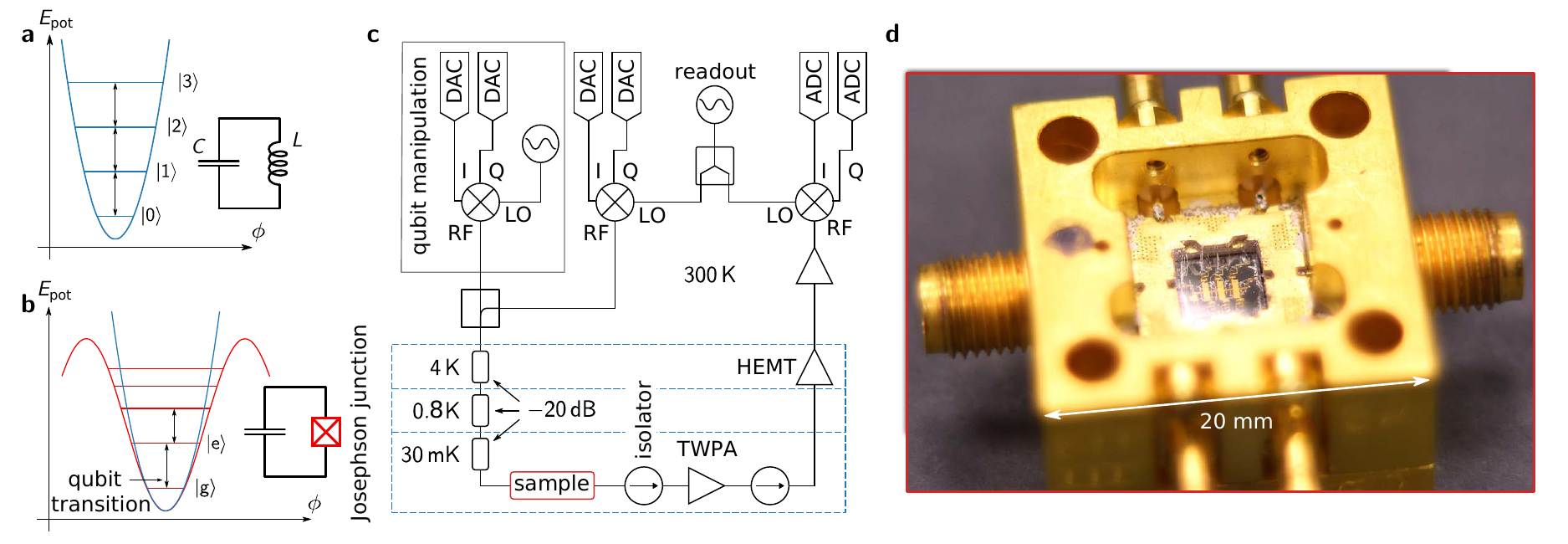}
\caption{(a) A simple harmonic resonator formed by a capacitor $C$ and an inductor $L$. The potential energy is the energy stored in the inductor and assumes a parabolic shape with respect to the phase variable $\phi$, related to the flux induced by the inductor. The energy levels, called Fock states, are equidistant, such that transitions cannot be addressed individually. (b) By replacing the inductor with a Josephson junction, the potential of the transmon becomes anharmonic, which isolates two energy levels to form a computational qubit basis. (c) Simplified schematic microwave setup, including pulse generation and processing. Microwave pulses are generated by microwave sources and arbitrary waveform generators. 
IQ-mixers facilitate phase sensitive amplitude modulation. (d) Photograph of a fabricated sample mounted in a gold-plated copper package.}
\label{fig:fig1}
\end{figure*}

The building blocks of superconducting quantum computing hardware are superconducting qubits, solid-state artificial atoms with level transitions in the microwave regime \cite{Clarke2008}. The transmon qubit~\cite{Koch2007} has emerged as one of the most popular qubit designs due to its robust fabrication process, demonstrated operation and readout, and reproducible lifetimes and coherence times in the order of tens of microsenconds~\cite{Oliver2013}. It is closely related to a harmonic $LC$-oscillator, which features equidistant energy levels, illustrated in Fig.~\ref{fig:fig1}(a). Coherent control requires an isolated pair of energy levels that form a computational qubit basis \cite{Schoelkopf2008}, and this motivates the need for anharmonic oscillators. The necessary anharmonicity is provided by the Josephson junction\hspace{0.03cm}---\hspace{0.03cm}a lithographically defined tunnel barrier between two superconducting electrodes\hspace{0.03cm}---\hspace{0.03cm}that behaves as a non-linear inductor without any significant dissipation \cite{Oliver2013}. The schematic transmon circuit is depicted in Fig.~\ref{fig:fig1}(b). 

By varying the relative strengths of the energies associated with the inductance, capacitance, and tunnel elements in the circuit, various architectures of superconducting qubits can be realized \cite{Devoret2013}, each featuring their own unique noise susceptibility and operation regime \cite{Clarke2008}.

The computational qubit basis is spanned by the ground and excited states $\Ket g$ and $\Ket e$, very much like the two states of a classical bit. However, quantum mechanics enables a qubit to be in any superposition state $\psi=\alpha\Ket g +\beta\Ket e$ with probability amplitudes $\alpha$ and $\beta$. Due to quantum correlations (another fundamental concept referred to as entanglement), this leads to quantum parallelism and quantum interference, the two fundamental principles substantiating the power of quantum computations.

\subsection{Microwave Regime}

Despite their macroscopic size, superconducting circuits behave quantum coherently when cooled to milli-Kelvin temperatures. This is mainly due to the absence of conductivity losses in the superconductor when cooled below its critical temperature. Superconducting circuits are fabricated with elementary superconductors such as aluminum, niobium or related compounds such as \ch{NbN} with critical temperatures between $\SI{1}{K}$ and $\SI{16}{K}$. The circuit operation temperature is small compared to the superconducting gap ($\geq\SI{50}{GHz}$) which further suppresses the losses induced by residual unpaired electrons (quasi-particles). The sample operation temperature is $T\sim\SI{10}{mK}$, achieved by $^4$\ch{He}/$^3$\ch{He}-dilution refrigerators, and corresponds to a frequency of $f=k_{\mathrm{B}} T/h \sim \SI{0.2}{GHz}$ (where $k_{\mathrm{B}}$ and $h$ are the Boltzmann and Planck constant, respectively), such that frequency transitions in the $\SI{5}{GHz}$ regime are only weakly thermally populated and the circuit can approximately be considered to remain in its ground-state in the absence of any controls. 

\subsection{Readout and Control of Superconducting Qubits}

Today, superconducting qubit state measurements are most commonly performed using a dispersive readout scheme \cite{Blais2004,Wallraff2004}. The readout device is a resonator that is weakly coupled to the qubit at a detuned frequency. Due to a qubit state dependent ``dressing'' of the readout resonator, the qubit state can be inferred by spectroscopically probing a dispersive shift of $\sim\SI{1}{MHz}$ in its resonance frequency. This scheme enables a quantum non-demolition measurement, where the qubit is mapped onto one of its basis states that corresponds to the measurement outcome \cite{Oliver2013}.

After applying a sequence of quantum gate operations or allowing a free qubit evolution, a measurement process is initiated by populating the readout resonator with a microwave pulse. This is achieved by using an on-chip transmission line coupled to the readout resonator which enables the measurement of microwave reflection or transmission.

Qubit excitation and quantum gates are performed by applying microwave drive pulses at or close to the qubit transition frequency. The qubit undergoes coherent oscillations between its two fundamental basis states, referred to as Rabi oscillations, which can be stopped at any point in time to prepare a desired superposition state.
Microwave pulses inducing single-qubit rotations 
have a typical duration of $\sim\SI{20}{\nano \second}$, and are amplitude modulated by a Gaussian envelope to achieve a localized pulse in Fourier space.

\subsection{Experimental Setup}\label{sec:setup}
\begin{figure*}[ht!]
\centering
\includegraphics[width=0.925\textwidth]{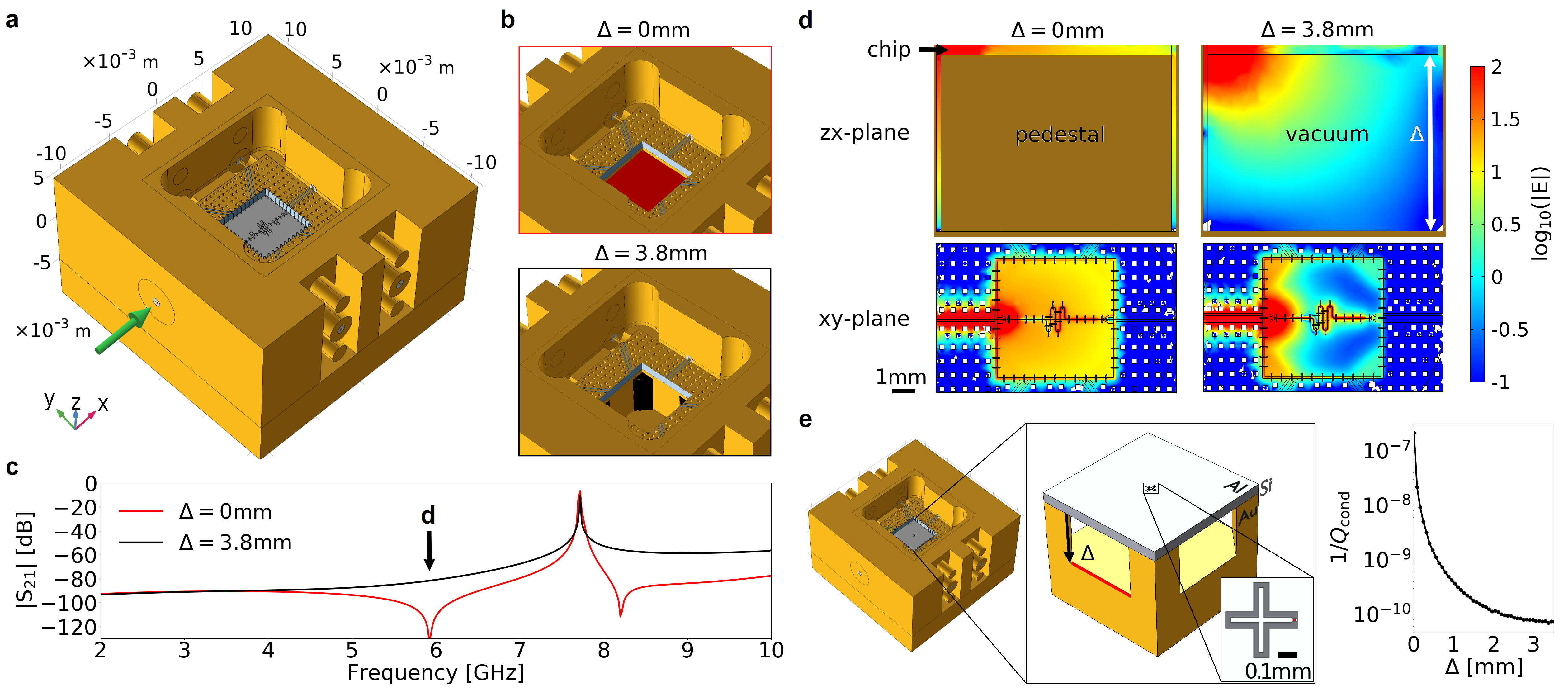}
\caption{Microwave simulations of the sample package shown in Fig.~\ref{fig:fig1}(d). (a) Gold-plated copper package (without lid) with wire-bonded superconducting circuit chip. The green arrow indicates the microwave input port. (b) Comparison of the transmission spectra and conductivity loss with a solid pedestal ($\Delta = \SI{0}{\milli \meter}$, colored in red), and a drilled out pedestal with four corner posts ($\Delta = \SI{3.8}{\milli \meter}$, black). (c) Simulated transmission magnitude spectrum $\vert S_{21}\vert$ of the chip holding an interrupted transmission line resonator with a resonance frequency at $\sim\SI{7.7}{GHz}$ mounted inside the package with pedestal (red) and a drilled out pedestal (black). The package with the drilled out pedestal is free of any undesired package modes in the band of interest. (d) E-field magnitude ($\SI{}{V/\meter}$) plots in the zx- and xy-plane at $\SI{5.9}{GHz}$ (indicated with an arrow in (c)). The presence of the pedestal enhances the E-field magnitude in the chip and direct vicinity. (e) Simulated conductivity loss $1/Q_{\rm cond}$ due to the normal conducting ($\sigma=\SI{4.5e9}{S/\meter}$ \cite{Lide2005}) package, extracted from a transmon qubit (approximated as linear resonator) in the center. The conductivity loss depends on the gap size $\Delta$ between the chip and the pedestal (the four corner posts remain in place).}
\label{fig:fig2}
\end{figure*}

A typical microwave measurement setup is schematically depicted in Fig.~\ref{fig:fig1}(c). 
Room temperature electronics feed microwave pulses to the cryostat through coaxial cables with a characteristic impedance matched to $\SI{50}{$\Omega$}$. The signals 
pass a series of attenuators which are thermally anchored to the different temperature stages, respectively, in order to sequentially reduce the room temperature Johnson-Nyquist noise, which forms a decoherence channel for qubits. 

Measurement pulses that are transmitted through the sample are passed through microwave isolators before they reach a quantum limited amplifier such as a travelling wave parametric amplifier (TWPA) \cite{Macklin307}. The TWPA amplifies microwave signals of individual photons by about $\SI{20}{dB}$ in a broad band of $\sim\SI{2}{GHz}$. It works close to a regime where only the minimum amount of noise dictated by quantum mechanics is added to the amplified signal, known as quantum limited amplification. Such sensitive amplification enables a single-shot readout of a set of qubits and facilitates the implementation of real-time quantum feedback \cite{Riste2013a}. The isolators prevent leakage of the pump tone, required to operate the TWPA, back to the sample. After passing a low-noise high electron mobility amplifier (HEMT) at the outputs, thermalized to $\SI{4}{K}$, the signal is further amplified at room temperature before being processed in room temperature microwave electronics.

\subsection{Sample Design and Fabrication Techniques}

Superconducting quantum circuits are fabricated using commercially available low-loss silicon or sapphire substrates. Primary fabrication processes include thin film evaporation or sputtering of superconducting materials, structured by optical or electron-beam lithography.
The 
dielectric for the Josephson junction tunnel barrier is formed by a controlled in-situ oxidation. 
The properties and mutual couplings of circuit elements can be individually tailored and fabricated in a reproducible manner due to their macroscopic physical sizes (millimeter-scale) \cite{Schoelkopf2008}.


\section{Microwave Package Engineering}\label{sec:uwcontrol}
Superconducting quantum chips are mounted into a sample package which is thermally attached to the cold stage of a dilution refrigerator. It defines the immediate electromagnetic environment of the qubits and connects the quantum circuit to the coaxial control lines. 

The primary purpose of a package is to simultaneously shield the quantum circuit from the environment while enabling its efficient control and thermalization.
The package can either act as a 3D cavity with an engineered mode spectrum and a high-quality resonance mode used for qubit readout \cite{Paik2011}, or merely provide an electromagnetic environment with suppressed spurious modes in the frequency spectrum of interest.
The material and geometric design of the sample package need to be chosen such that qubit energy loss channels are suppressed. The qubit can spontaneously dissipate energy to dielectric defects on the surface and interfaces of the sample or by coupling to unwanted package modes. Furthermore, normal metals introduce conductivity loss, which grows with increasing electrical resistance. While conductivity loss in superconducting packages is suppressed, their thermal conductivity is in general strongly reduced, which can result in improper sample thermalization.

Typical package materials are aluminum ($\sigma_{\rm Al}$: superconducting; $\kappa_{\rm Al}$: limited; oxide), copper ($\sigma_{\rm Cu}$: high; $\kappa_{\rm Cu}$: high; oxide), or gold-plated copper ($\sigma_{\rm Au-Cu}$: high; $\kappa_{\rm Au-Cu}$: high; limited oxide). The relations between electrical conductivity $\sigma$ and thermal conductivity $\kappa$ are $\sigma_{\rm Al}\gg\sigma_{\rm Cu} \gtrsim \sigma_{\rm Au-Cu}$ and $\kappa_{\rm Al}\ll\kappa_{\rm Au-Cu} \lesssim \kappa_{\rm Cu}$. Fig.~\ref{fig:fig1}(d) shows the gold-plated copper package used in the simulations presented here. The copper core ensures high thermal conductivity whereas the gold-plating suppresses the formation of a surface oxide layer.

Metallic waveguides for quantum circuit control are imprinted on a dielectric circuit board (interposer \cite{Bronn2018}) and matched to a characteristic impedance of $\SI{50}{$\Omega$}$. The gold-plated interposer\hspace{0.03cm}---\hspace{0.03cm}made from Rogers TMM\textregistered~10 ceramics\hspace{0.03cm}---\hspace{0.03cm}routes the coaxial control lines to the quantum circuit connected through aluminum wire bonds.
Vertical vias in the interposer reduce cross-talk between different coplanar signal lines. 

The 3D finite element simulation software {COMSOL Multiphysics\textregistered} is used to analyze the microwave properties of the package shown in Fig.~\ref{fig:fig1}(d). The package model is schematically depicted in Fig.~\ref{fig:fig2}(a), holding the interposer and a chip with an interrupted transmission line resonator. Ideally, one expects a pronounced peak at the resonance frequency of this resonator at $\sim\SI{7.7}{GHz}$ and no transmission away from resonance. When measured inside the sample package, the transmission spectrum of the resonator chip is convoluted with a broad package mode centered at $\sim\SI{19}{GHz}$, mediating a non-zero baseline transmission.

Fig.~\ref{fig:fig2}(b) shows schematic drawings of two sample package versions, one with the chip sitting on a solid pedestal (red), and one where the chip is supported by four corner posts (black) with the pedestal drilled out to a distance of $\Delta=\SI{3.8}{\milli \meter}$ \cite{Bronn2018,Wenner2011}. The response of the on-chip resonator is visible in both simulated transmission magnitude spectra $|S_{21}|$ (red and black lines in Fig.~\ref{fig:fig2}(c)). The simulation of the package with pedestal (red) reveals pronounced package modes in the relevant frequency range which provides a potential qubit loss channel. They are suppressed in the version (black) with the pedestal drilled out, indicating an electromagnetic sample environment without spurious modes in the frequency range of interest. Fig.~\ref{fig:fig2}(d) shows the electric field magnitude distribution at one of the box modes at $\SI{5.9}{GHz}$. For the package with solid pedestal, the electric field is strongly enhanced inside and in the vicinity of the chip.  

Normal metal near the sample forms a loss channel due to finite conductivity. The conductivity loss $Q_{\mathrm{cond}}^{-1}$ is simulated using a surface participation model. The simulation estimates the conductivity loss of a typically sized transmon qubit with a ``+''-shaped capacitor \cite{Barends2013}, defined on a chip that is mounted in the presented package, as a function of the distance $\Delta$ between sample and pedestal, see Fig.~\ref{fig:fig2}(e). While the conductivity loss assumes a maximum for $\Delta=\SI{0}{mm}$, it saturates at $\Delta\sim\SI{3}{mm}$ and is suppressed by about three orders of magnitude. The extracted conductivity loss for a package with solid pedestal is $Q_{\rm cond}=\SI{4.5e6}{}$. Some of the highest experimentally achieved qubit lifetimes of $T_1\approx\SI{150}{\micro s}$ \cite{Oliver2013} correspond to a quality factor of $Q\approx\SI{4.5e6}{}$ implying that conductivity loss cannot be neglected for long-lived qubits.




\section{Conclusion}

Superconducting circuits are operated in the microwave regime, which enables a high degree of control and provides a rich toolbox of experimental techniques. In return, quantum circuits likewise couple to unwanted microwave modes, making it necessary to provide an engineered electromagnetic environment free of spurious modes in the frequency range of interest, ensured by the sample package. Microwave simulations of the presented sample package indicate that a solid support pedestal below the sample introduces spurious package modes which form a decoherence channel for qubits. In addition, simulations show that the conductivity loss due to a solid pedestal limits the lifetime of long-lived qubits. The conductivity loss can be mitigated by about three orders of magnitude in the studied package by removing the support pedestal.



\section*{Acknowledgment}
This research is funded by the Office of the Director of National Intelligence (ODNI), Intelligence Advanced Research Projects Activity (IARPA), and by the Assistant Secretary of Defense for Research \& Engineering under Air Force Contract No. FA8721-05-C-0002.





\bibliographystyle{IEEEtran}
\bibliography{IEEE_ben_only,IEEE_jochen}
\end{document}

%% file: IMS_modify_IEEEtran_18b_CTAN_V3c.tex
\makeatletter

\def\@IMSauthorblockNAMEstyle{\normalfont\IMSauthorsize}
\def\@IMSauthorblockAFFILstyle{\normalfont\IMSaffilsize}
\def\@IMSauthorblockEMAILstyle{\normalfont\IMSaffilsize}
\def\IMSauthorblockNAME#1{%
\relax\@IMSauthorblockNAMEstyle%
#1%
}%
\def\IMSauthorblockAFFIL#1{%
\relax\@IMSauthorblockAFFILstyle%
\vskip\@IEEEauthorblockAtopspace
#1%
}%
\def\IMSauthorblockEMAIL#1{%
\relax\@IMSauthorblockEMAILstyle%
\vskip\@IEEEauthorblockAtopspace
#1%
}%
\newcommand{\IMSauthor}[1]{%
\ifIsBlindReviewVersion%
\author{\phantom{\parbox{\textwidth}{\center\relax#1}}}%
\else%
\author{\parbox{\textwidth}{\center\relax#1}}%
\fi%
}%
\newif\ifIsBlindReviewVersion
\def\IMSthispaperforblindreview{\IsBlindReviewVersiontrue}
\def\IMSthispaperforfinalpublication{\IsBlindReviewVersionfalse}
\def\@maketitle{\newpage
\bgroup\par\addvspace{0.5\baselineskip}\centering%
\ifCLASSOPTIONtechnote
   {\bfseries\large\@IEEEcompsoconly{\sffamily}\@title\par}\vskip 1.3em{\lineskip .5em\@IEEEcompsoconly{\sffamily}\@author
   \@IEEEspecialpapernotice\par{\@IEEEcompsoconly{\vskip 1.5em\relax
   \@IEEEtitleabstractindextextbox{\@IEEEtitleabstractindextext}\par
   \hfill\@IEEEcompsocdiamondline\hfill\hbox{}\par}}}\relax
\else
   \vskip0.2em{\IMStitlesize\ifCLASSOPTIONtransmag\bfseries\LARGE\fi\@IEEEcompsoconly{\sffamily}\@IEEEcompsocconfonly{\normalfont\normalsize\vskip 2\@IEEEnormalsizeunitybaselineskip
   \bfseries\Large}\@title\par}\vskip1.0em\par
   \ifCLASSOPTIONconference%
      {\@IEEEspecialpapernotice\mbox{}\vskip\@IEEEauthorblockconfadjspace%
       \mbox{}\hfill\begin{@IEEEauthorhalign}\@author\end{@IEEEauthorhalign}\hfill\mbox{}\par}\relax
   \else
      \ifCLASSOPTIONpeerreviewca
         {\@IEEEcompsoconly{\sffamily}\@IEEEspecialpapernotice\mbox{}\vskip\@IEEEauthorblockconfadjspace%
          \mbox{}\hfill\begin{@IEEEauthorhalign}\@author\end{@IEEEauthorhalign}\hfill\mbox{}\par
          {\@IEEEcompsoconly{\vskip 1.5em\relax
           \@IEEEtitleabstractindextextbox{\@IEEEtitleabstractindextext}\par\hfill
           \@IEEEcompsocdiamondline\hfill\hbox{}\par}}}\relax
      \else
         \ifCLASSOPTIONtransmag
           {\@IEEEspecialpapernotice\mbox{}\vskip\@IEEEauthorblockconfadjspace%
            \mbox{}\hfill\begin{@IEEEauthorhalign}\@author\end{@IEEEauthorhalign}\hfill\mbox{}\par
           {\vspace{0.5\baselineskip}\relax\@IEEEtitleabstractindextextbox{\@IEEEtitleabstractindextext}\vspace{-1\baselineskip}\par}}\relax
         \else
           {\lineskip.5em\@IEEEcompsoconly{\sffamily}\sublargesize\@author\@IEEEspecialpapernotice\par
           {\@IEEEcompsoconly{\vskip 1.5em\relax
            \@IEEEtitleabstractindextextbox{\@IEEEtitleabstractindextext}\par\hfill
            \@IEEEcompsocdiamondline\hfill\hbox{}\par}}}\relax
         \fi
      \fi
   \fi
\fi\par\addvspace{0.0\baselineskip}\egroup}

\def\IMStitlesize{\@setfontsize{\IMStitlesize}{18}{21pt}}
\def\IMSauthorsize{\@setfontsize{\IMSauthorsize}{12}{13pt}}
\def\IMSaffilsize{\@setfontsize{\IMSaffilsize}{12}{13pt}}
\def\IMScaptionsize{\@setfontsize{\IMScaptionsize}{8}{9pt}}
\def\IMSbibsize{\@setfontsize{\IMSbibsize}{8}{9pt}}

\def\@IEEEauthorblockNstyle{\IMSauthorsize\@IEEEcompsocnotconfonly{\sffamily}\@IEEEcompsocconfonly{\large}}
\def\@IEEEauthorblockAstyle{\IMSaffilsize\@IEEEcompsocnotconfonly{\sffamily}\@IEEEcompsocconfonly{\itshape}\@IEEEcompsocconfonly{\large}}
\def\@IEEEauthordefaulttextstyle{\IMSauthorsize\@IEEEcompsocnotconfonly{\sffamily}\sublargesize}

\def\thebibliography#1{\section*{\refname}%
    \addcontentsline{toc}{section}{\refname}%
    \IMSbibsize\@IEEEcompsocconfonly{\small}\vskip 0.3\baselineskip plus 0.1\baselineskip minus 0.1\baselineskip
    \list{\@biblabel{\@arabic\c@enumiv}}%
    {\settowidth\labelwidth{\@biblabel{#1}}%
    \leftmargin\labelwidth
    \advance\leftmargin\labelsep\relax
    \itemsep \IEEEbibitemsep\relax
    \usecounter{enumiv}%
    \let\p@enumiv\@empty
    \renewcommand\theenumiv{\@arabic\c@enumiv}}%
    \let\@IEEElatexbibitem\bibitem%
    \def\bibitem{\@IEEEbibitemprefix\@IEEElatexbibitem}%
\def\newblock{\hskip .11em plus .33em minus .07em}%
\ifCLASSOPTIONtechnote\sloppy\clubpenalty4000\widowpenalty4000\interlinepenalty100%
\else\sloppy\clubpenalty4000\widowpenalty4000\interlinepenalty500\fi%
    \sfcode`\.=1000\relax}

\long\def\@makecaption#1#2{%
\ifx\@captype\@IEEEtablestring%
\par\@IEEEtabletopskipstrut
\else
\@IEEEfigurecaptionsepspace
\fi
\setbox\@tempboxa\hbox{\normalfont\IMScaptionsize {#1.}\nobreakspace\nobreakspace #2}%
\ifdim \wd\@tempboxa >\hsize%
\setbox\@tempboxa\hbox{\normalfont\IMScaptionsize {#1.}\nobreakspace\nobreakspace}%
\parbox[t]{\hsize}{\normalfont\IMScaptionsize\noindent\unhbox\@tempboxa#2}%
\else
\ifCLASSOPTIONconference \hbox to\hsize{\normalfont\IMScaptionsize\hfil\box\@tempboxa\hfil}%
\else \hbox to\hsize{\normalfont\IMScaptionsize\box\@tempboxa\hfil}%
\fi\fi
\ifx\@captype\@IEEEtablestring%
\@IEEEtablecaptionsepspace
\else
\fi}

\newlength\tablecaptiontotableskip
\newlength\figuretocaptionskip
\def\@IEEEfigurecaptionsepspace{\vskip\figuretocaptionskip\relax}%
\def\@IEEEtablecaptionsepspace{\vskip\tablecaptiontotableskip\relax}%

\def\abstract{\normalfont%
\@IEEEabskeysecsize\bfseries\textit{\abstractname}\,\bfseries\textit{---}\,%
\@IEEEgobbleleadPARNLSP}%

\def\IEEEkeywords{\normalfont%
\@IEEEabskeysecsize\bfseries\textit{\IEEEkeywordsname}\,\bfseries\textit{---}\,%
\@IEEEgobbleleadPARNLSP}%
\def\endIEEEkeywords{\relax\vspace{0.67ex}%
\par\if@twocolumn\else\endquotation\fi%
\normalsize\normalfont}%

\DeclareRobustCommand*{\IMSauthorrefmark}[1]{\raisebox{0pt}[0pt][0pt]{\textsuperscript{\footnotesize{#1}}}}%
\def\@IEEEauthorblockNtopspace{0ex}
\def\@IEEEauthorblockAtopspace{1mm}
\def\IEEEkeywordsname{Keywords}
\def\subsubsection{\@startsection{subsubsection}{3}{\z@}{1.5ex plus 1.5ex minus 0.5ex}%
{0.7ex plus .5ex minus 0ex}{\normalfont\normalsize\itshape}}%
\newlength{\CPheadmatchindent}%
\def\@seccntformat#1{\hbox to\CPheadmatchindent{\csname the#1dis\endcsname}\hskip 0.1em \relax}
\IEEEilabelindentA \parindent
\IEEEilabelindent \IEEEilabelindentA
\IEEEelabelindent \parindent
\IEEEdlabelindent \parindent
\IEEElabelindent \parindent

\newlength\@IMSparindent
\newcommand\IMSdisplayacksection[1]{%
\ifIsBlindReviewVersion%
\noindent\phantom{\parbox[t]{\columnwidth}{\normalbaselines\setlength{\parindent}{\@IMSparindent}{#1}\strut}}
\else%
\noindent\parbox[t]{\columnwidth}{\normalbaselines\setlength{\parindent}{\@IMSparindent}{#1}\strut}%
\fi%
}%

\makeatother

%% file: IEEE_Microwave_QC_Vertical.bbl
\begin{thebibliography}{10}
\providecommand{\url}[1]{#1}
\csname url@samestyle\endcsname
\providecommand{\newblock}{\relax}
\providecommand{\bibinfo}[2]{#2}
\providecommand{\BIBentrySTDinterwordspacing}{\spaceskip=0pt\relax}
\providecommand{\BIBentryALTinterwordstretchfactor}{4}
\providecommand{\BIBentryALTinterwordspacing}{\spaceskip=\fontdimen2\font plus
\BIBentryALTinterwordstretchfactor\fontdimen3\font minus
  \fontdimen4\font\relax}
\providecommand{\BIBforeignlanguage}[2]{{%
\expandafter\ifx\csname l@#1\endcsname\relax
\typeout{** WARNING: IEEEtran.bst: No hyphenation pattern has been}%
\typeout{** loaded for the language `#1'. Using the pattern for}%
\typeout{** the default language instead.}%
\else
\language=\csname l@#1\endcsname
\fi
#2}}
\providecommand{\BIBdecl}{\relax}
\BIBdecl

\bibitem{Shor1994}
P.~W. Shor, ``Algorithms for quantum computation: discrete logarithms and
  factoring,'' in \emph{Proceedings 35th Annual Symposium on Foundations of
  Computer Science}, Nov 1994, pp. 124--134.

\bibitem{Grover1996}
L.~K. Grover, ``A fast quantum mechanical algorithm for database search,''
  \emph{Proceedings, 28th Annual ACM Symposium on the Theory of Computing}, p.
  212, may 1996.

\bibitem{Oliver2013}
W.~D. Oliver and P.~B. Welander, ``Materials in superconducting qubits,''
  \emph{MRS Bulletin}, vol.~38, pp. 816--825, oct 2013.

\bibitem{Steffen2006}
M.~Steffen, M.~Ansmann, R.~C., N.~Bialczak, E.~L. Katz, R.~McDermott,
  M.~Neeley, E.~M. Weig, A.~N. Cleland, and J.~M. Martinis, ``Measurement of
  the entanglement of two superconducting qubits via state tomography,''
  \emph{Science}, vol. 313, no. 5792, pp. 1423--1425, 2006.

\bibitem{Barends2014}
R.~Barends, J.~Kelly, A.~Megrant, A.~Veitia, D.~Sank, E.~Jeffrey, T.~C. White,
  J.~Mutus, A.~G. Fowler, B.~Campbell, Y.~Chen, B.~Chiaro, A.~Dunsworth,
  C.~Neill, P.~O'Malley, P.~Roushan, A.~Vainsencher, J.~Wenner, A.~N. Korotkov,
  A.~N. Cleland, and J.~M. Martinis, ``Superconducting quantum circuits at the
  surface code threshold for fault tolerance,'' \emph{Nature}, vol. 508, no.
  7497, pp. 500--503, 2014.

\bibitem{Gustavsson2013}
S.~Gustavsson, O.~Zwier, J.~Bylander, F.~Yan, F.~Yoshihara, Y.~Nakamura, T.~P.
  Orlando, and W.~D. Oliver, ``Improving quantum gate fidelities by using a
  qubit to measure microwave pulse distortions,'' \emph{Phys. Rev. Lett.}, vol.
  110, no.~4, jan 2013.

\bibitem{Blais2004}
A.~Blais, R.-S. Huang, A.~Wallraff, S.~M. Girvin, and R.~J. Schoelkopf,
  ``Cavity quantum electrodynamics for superconducting electrical circuits: An
  architecture for quantum computation,'' \emph{Phys. Rev. A}, vol.~69, no.~6,
  p. 062320, Jun 2004.

\bibitem{Wallraff2004}
A.~Wallraff, D.~I. Schuster, A.~Blais, L.~Frunzio, R.-S. Huang, J.~Majer,
  S.~Kumar, S.~M. Girvin, and R.~J. Schoelkopf, ``Strong coupling of a single
  photon to a superconducting qubit using circuit quantum electrodynamics,''
  \emph{Nature}, vol. 431, no. 7005, pp. 162--167, Sep 2004.

\bibitem{Lucero2012}
E.~Lucero, R.~Barends, Y.~Chen, J.~Kelly, M.~Mariantoni, A.~Megrant,
  P.~{O'Malley}, D.~Sank, A.~Vainsencher, J.~Wenner, T.~White, Y.~Yin, A.~N.
  Cleland, and J.~M. Martinis, ``Computing prime factors with a josephson phase
  qubit quantum processor,'' \emph{Nat. Phys.}, vol.~8, no.~10, pp. 719--723,
  Oct. 2012.

\bibitem{Colless2018}
J.~I. Colless, V.~V. Ramasesh, D.~Dahlen, M.~S. Blok, M.~E. Kimchi-Schwartz,
  J.~R. McClean, J.~Carter, W.~A. de~Jong, and I.~Siddiqi, ``Computation of
  molecular spectra on a quantum processor with an error-resilient algorithm,''
  \emph{Phys. Rev. X}, vol.~8, p. 011021, 2018.

\bibitem{Clarke2008}
J.~Clarke and F.~K. Wilhelm, ``Superconducting quantum bits,'' \emph{Nature},
  vol. 453, 2008.

\bibitem{Koch2007}
J.~Koch, T.~M. Yu, J.~Gambetta, A.~A. Houck, D.~I. Schuster, J.~Majer,
  A.~Blais, M.~H. Devoret, S.~M. Girvin, and R.~J. Schoelkopf,
  ``Charge-insensitive qubit design derived from the cooper pair box,''
  \emph{Phys. Rev. A}, vol.~76, p. 042319, Oct 2007.

\bibitem{Schoelkopf2008}
R.~J. Schoelkopf and S.~M. Girvin, ``Wiring up quantum systems,''
  \emph{Nature}, vol. 451, pp. 664--669, 2008.

\bibitem{Devoret2013}
M.~H. Devoret and R.~J. Schoelkopf, ``Superconducting ciruits for quantum
  information: an outlook,'' \emph{Science}, vol. 339, no. 6124, pp.
  1169--1174, 2013.

\bibitem{Lide2005}
D.~R. Lide, ``Crc handbook of chemistry and physics, internet version 2005,''
  \emph{CRC Press, Boca Raton, FL}, 2005.

\bibitem{Macklin307}
C.~Macklin, K.~O{\textquoteright}Brien, D.~Hover, M.~E. Schwartz,
  V.~Bolkhovsky, X.~Zhang, W.~D. Oliver, and I.~Siddiqi, ``A
  near{\textendash}quantum-limited josephson traveling-wave parametric
  amplifier,'' \emph{Science}, vol. 350, no. 6258, pp. 307--310, 2015.

\bibitem{Riste2013a}
D.~Rist\`e, M.~Dukalski, C.~A. Watson, G.~de~Lange, M.~J. Tiggelman, Y.~M.
  Blanter, K.~W. Lehnert, R.~N. Schouten, and L.~DiCarlo, ``Deterministic
  entanglement of superconducting qubits by parity measurement and feedback,''
  \emph{Nature}, vol. 502, p. 350, Oct. 2013.

\bibitem{Paik2011}
H.~Paik, D.~I. Schuster, L.~S. Bishop, G.~Kirchmair, G.~Catelani, A.~P. Sears,
  B.~R. Johnson, M.~J. Reagor, L.~Frunzio, L.~I. Glazman, S.~M. Girvin, M.~H.
  Devoret, and R.~J. Schoelkopf, ``Observation of high coherence in {Josephson}
  junction qubits measured in a three-dimensional circuit {QED} architecture,''
  \emph{Phys. Rev. Lett.}, vol. 107, no.~24, 2011.

\bibitem{Bronn2018}
N.~T. Bronn, V.~P. Adiga, S.~B. Olivadese, X.~Wu, J.~M. Chow, and D.~P. Pappas,
  ``High coherence plane breaking packaging for superconducting qubits,''
  \emph{Quantum Science and Technology}, vol.~3, no.~2, feb 2018.

\bibitem{Wenner2011}
J.~Wenner, M.~Neeley, R.~C. Bialczak, M.~Lenander, E.~Lucero, A.~D.
  O’Connell, D.~Sank, H.~Wang, M.~Weides, A.~N. Cleland, and J.~M. Martinis,
  ``Wirebond crosstalk and cavity modes in large chip mounts for
  superconducting qubits,'' \emph{Supercond. Sci. Technol}, vol.~24, p. 065001,
  2011.

\bibitem{Barends2013}
R.~Barends, J.~Kelly, A.~Megrant, D.~Sank, E.~Jeffrey, Y.~Chen, Y.~Yin,
  B.~Chiaro, J.~Mutus, C.~Neill, P.~O'Malley, P.~Roushan, J.~Wenner, T.~White,
  A.~N. Cleland, and J.~M. Martinis, ``Coherent {Josephson} qubit suitable for
  scalable quantum integrated circuits,'' \emph{Phys. Rev. Lett.}, vol. 111,
  no.~8, p. 080502, 2013.

\end{thebibliography}
